\documentclass[12pt]{article}
\usepackage{latexsym}
\newcommand{\beq}{\begin{eqnarray}}
\newcommand{\eeq}{\end{eqnarray}}
\newcommand{\snp}{{\sum_N}'}
\newcommand{\nn}{\nonumber}
\newcommand{\ina}{\frac 1 {2\pi i} \int\limits_{c-i\infty}^{c+i\infty}}
\newcommand{\ing}{\ina d\alpha \,\,\Gamma (\alpha)}
\newcommand{\en}{\left( E_N -E_0 \right)}
\newcommand{\muc}{\left(e^{-\beta (\mu_c -\mu)}\right)}
\newcommand{\pa}{\partial}
\newcommand{\cam}{{\cal M}}
\newcommand{\reals}{\mbox{${\rm I\!R }$}}

\begin{document}
\title{Bose-Einstein condensation under external conditions}
\author{Klaus Kirsten    \cite{kk}\\University of Leipzig\\
Institute of Theoretical Physics\\
Augustusplatz 10, 
04109 Leipzig\\
\\and\\
\\
David J. Toms   \cite{djt}          \\
Department of Physics\\ University of Newcastle Upon Tyne,\\
Newcastle Upon Tyne, U. K. NE1 7RU}

\date{\today}
\maketitle

\begin{abstract}
We discuss the phenomenon of Bose-Einstein condensation under general
external conditions using connections between partition sums and the 
heat-equation. Thermodynamical quantities like the critical
temperature are given in terms of the heat-kernel coefficients of the 
associated Schr\"odinger equation. The general approach is applied to 
situations where the gas is confined by arbitrary potentials 
or by boxes of arbitrary shape. 
\end{abstract}
\eject
One of the most striking developments of the last few years has been the 
experimental observation of Bose-Einstein condensation in very cold
gases of rubidium \cite{rub}, lithium \cite{lith} 
and sodium \cite{sod}. The systems 
are very dilute and as a first approximation would be expected to 
be well described by a boson gas model with no interactions among 
the atoms. The atoms are confined in complicated magnetic traps 
which can be modeled by harmonic oscillator potentials. Nowadays 
one can say that the thermodynamic properties of the confined gas 
are very well understood, including the effects that the finite 
number of particles has on the system 
\cite{pra,martin,finiten,ravndal,brosens,gerd} as well as the 
influence of a weak interaction among the atoms \cite{weakint}. The 
above mentioned results have all been obtained in the grand  
canonical approach, which, although inadequate for the calculation 
of, for example, flucuations of particle numbers 
\cite{mikro1,mikro2,mikro3,mikro4}
is still reliable for 
thermodynamical properties such as the critical temperature or the 
ground state occupation number. 

Although recently most interest is in Bose-Einstein condensation in
magnetic traps also different confining potentials \cite{MIT} or 
gases confined inside a rectangular box \cite{box} have been considered and 
specific differences have been stated. 
A formal discussion of boundary effectis has been given in \cite{ziff}.
All this research 
may be called analysis of Bose-Einstein condensation under external 
conditions. 

In this article we want to provide an approach which 
enables one to deal with any kind of external conditions (including 
the most relevant anisotropic harmonic oscillator potential). The methods 
we are going to employ are the so-called heat-kernel techniques 
extensively used in finite temperature relativistic quantum field theory
starting with the work of Dowker and Kennedy \cite{11}. (See also 
\cite{12}.) In non-relativistic theories these techniques are however nearly 
unemployed and we want to show in the present article that they can also
be used here in a very effective way. 

Let us consider a system of $N$ non-interacting bosons under external 
conditions such that the single particle state energies $E_N$ are 
determined by the Schr\"odinger equation 
\beq
-\frac{\hbar ^2 }{2m} \Delta \phi_N (x) +
V(x) \phi_N (x) =E_N \phi_N (x)\;, \label{1}
\eeq
where $V(x)$ is the confining potential. For example in 
\cite{MIT} the confining potential 
\beq
V(x) = \sum_{i=1}^d \epsilon_i ^{p_i} |x_i|^{p_i} \label{1a}
\eeq
has been considered in the thermodynamic limit.
If in addition the gas is confined by a box ${\cal M}$ one has to 
impose suitable boundary conditions at the boundary $\partial{\cal M}$
of the box. Without specifying the spectrum $E_N$ (as a rule it will 
not be known explicitly) the grand canonical potential may be 
given in the form
\beq
q=q_0 +\sum_{n=1}^{\infty} \snp 
\frac 1 n e^{-\beta n (E_N -\mu) }\;.\label{2}
\eeq
Here 
\beq
q_0 = -d_0  \ln \left( 1-z e^{-\beta E_0 }\right)\;, \nn
\eeq
is the contribution of the ground state with energy $E_0$, $d_0$ 
is its degeneracy and $z=\exp (\beta \mu )$ is the fugacity. 
The prime in (\ref{2}) indicates that the ground state contribution is to be omitted from 
the sum. 

For the evaluation of this kind of expression it is very effective 
to make use of the Mellin-Barnes integral representation 
\cite{mellin}
\beq
e^{-v} = \ing v^{-\alpha } ,
\label{3}
\eeq
valid for $\Re v >0$ and $c\in \reals$ , $c>0$. This is easily proven 
by closing the contour to the left obtaining the power series expansion
of $\exp (-v) $. As an immediate consequence of eqs. (\ref{2}) 
and (\ref{3}) the grand canonical partition sum may be casted into the 
form 
\beq
q=q_0 +\ing \beta ^{-\alpha }
Li_{1+\alpha} \left(e^{-\beta (\mu_c -\mu)}\right)
 \zeta (\alpha), \label{4}  
\eeq
with the polylogarithm 
\beq
Li_n (x) =\sum_{l=1}^{\infty} \frac{x^l}{l^n} ,\label{5}
\eeq
and the spectral zeta function 
\beq
\zeta (\alpha) = \snp \en ^{-\alpha}.\label{6}
\eeq
Here we introduced the variable $\mu_c$ which denotes the value of the 
chemical potential at the transition temperature. For ideal gases this 
value is identical with the ground state energy, $\mu_c = E_0$.
The parameter $c$ in eq. (\ref{4}) is given in a way such that 
all poles of the integrand 
lie to the left of the contour. Closing the contour to the left 
one picks up the right-most residues of $\zeta (\alpha)$. Concerning their
determination there are deep connections between the zeta function of an
operator and its heat-kernel defined as 
\beq
K(t) = \snp e^{-t \en} .\label{7} 
\eeq
The small-t behaviour  
\beq
K(t) \sim \sum_{k=0}^{\infty} b_k t^{-j_k},\label{8}
\eeq
is of particular relevance here. $b_k$ are the so-called heat-kernel coefficients. We assume $j_k > j_{k+1}$,
and the connection between the residues of $\zeta(\alpha)$, which occur at $\alpha=j_k$, and the heat-kernel coefficients reads
\beq
{\rm Res } \,\,\zeta (\alpha=j_k) = \frac{b_k}{\Gamma (j_k)}.\label{9}
\eeq
Taking into consideration only the two right most poles of $\zeta (\alpha)$ one
arrives at 
\beq
q = q_0 + \beta^{-j_0} Li_{1+j_0} \muc b_0 + \beta^{-j_1} Li_{1+j_1} 
                       \muc b_1 +\ldots\;. \label{10}
\eeq
For the determination of the critical temperature and the thermodynamical 
properties of the gas up to $T_c$ the above approximation can be shown 
to be a very good one. 

We thus have a representation of the grand canonical partition sum in terms
of the heat-kernel coefficients of the relevant Schr\"odinger operator. 
What have we gained with this representation? This becomes clear by noticing
that the heat-kernel coeficients $b_0,b_1$ are known or easily determined
for a very large class of Schr\"odinger operators including the 
situation where the gas is confined by arbitrary potentials or boxes
of arbitrary shape. This includes physical situations which are 
of relevance
in recent experiments and also those which might be of relevance in future ones. 

Let us exemplify the power of the approach by considering further 
the particle number and the critical temperature. The particle number is 
\beq
N&=& \beta^{-1} \left(\frac{\pa q}{\pa \mu}\right) \left.\right|_{T,V}
\nn\\
&=& N_0 + \beta^{-j_0} Li_{j_0} \muc b_0 + \beta^{-j_1} Li_{j_1} 
                       \muc b_1 + ... \label{11}
\eeq
In order to avoid distinguishing all the different 
cases, let us assume from now on that $j_1 > 1$. In this case, the critical
temperature $1/\beta_c$ is determined by 
\beq
N= \beta_c ^{-j_0} \zeta_R (j_0) b_0 + \beta_c ^{-j_1} \zeta_R (j_1) b_1 +\ldots\;, \label{12} 
\eeq
and reads 
\beq
T_c = T_0 \left\{ 1-\frac{ \zeta_R (j_1) b_1 }{ j_0 \zeta_R (j_0) ^{j_1/j_0}
b_0 ^{j_1/j_0} } \frac 1 {N^{(j_0-j_1) / j_0 }} \right\} \label{13}
\eeq
with
\beq
T_0 = \frac 1 k 
\left( \frac N {\zeta_R (j_0) b_0 } \right) ^{1/j_0} \label{14}
\eeq
 the critical temperature in the bulk limit. Let us stress that eq. (\ref{13}) contains the 
influence that the finite number $N$ of particles have, the details
being encoded in the exponents $j_i$ and the coefficients
$b_i$. Proceeding with the energy and 
the specific heat, all thermodynamical quantities are easily displayed 
in terms of the heat-kernel coefficients. If $j_1 \leq 1$ different
results are found due to the behaviour of the polylogarithm 
\cite{inprep}. 

The presented technique allows us to calculate thermodynamical properties 
by treating the sums over the discrete energy levels. Another possible 
way to do this analysis is to approximate the sums by integrals. A 
crucial feature in obtaining a reliable approximation is to use an 
appropriate density of states \cite{martin}. By using a refinement 
of Karamatas' theorem \cite{Brownell}, one can show that the use 
of the density 
\beq
\rho (E ) = \frac{b_0}{\Gamma (j_0 ) } E^{j_0-1}
              +\frac{b_1}{\Gamma (j_1 ) } E^{j_1-1} \nn
\eeq
is completely equivalent to the analysis presented above.

A slightly different approach results by the use of the effective 
fugacity (for the anisotropic harmonic oscillator see \cite{ravndal})
\beq
z_{eff} = z e^{-\beta E_1} \nn
\eeq
where $E_1$ is the first excited level with, lets say, degeneracy $d_1$.
Whereas in the previous calculation only the groundstate has been treated 
separately, we now separate the groundstate and the first excited level
to find
\beq
q = q_0 +d_1 Li_1 (z_{eff}) + \beta ^ {-j_0} Li_{1+j_0} (z_{eff} ) b_0 
   + \beta ^ {-j_1} Li_{1+j_1} (z_{eff} ) b_1 + ...\nn
\eeq
This expansion is a very good approximation 
even below the critical temperature 
(for the harmonic oscillator see \cite{ravndal}).
Also here thermodynamical properties are obtained in the same manner as
before.

Let us now use our general results to consider the special situations 
mentioned already several times.

{\bf Example 1:} Gas inside a $d$-dimensional box of arbitrary shape.

In the notation of equation (\ref{1}) the situation is described by 
$V(x) =0$ and $\phi_N (x)|_{x\in \pa \cam} = 0 $. (Other boundary conditions
can be treated as well.) For 
this example the relevant heat-kernel coefficients are known
(see f.e. 
\cite{gil1}) and read
\beq
b_0 &=& (4\pi )^{-d/2} (2m/\hbar^2 ) ^{d/2} {\rm vol} (\cam ) \nn\\
b_1 &=& -(4\pi )^{-d/2} (2m/\hbar^2 ) ^{(d-1)/2} (\sqrt{\pi}/2) {\rm vol} (
\pa \cam ) \nn
\eeq
together with $j_0 = d/2$, $j_1 = (d-1)/2$. Here ${\rm vol} (\cam )$ and ${\rm vol} (\pa \cam )$ denote the finite volume of the box and its boundary respectively. For the critical temperature 
this means 
\beq
T_c = T_0 \left\{ 1+\frac 1 {2d} \frac{ {\rm vol} (\pa \cam )}
{({\rm vol} (\cam ))^{(d-1)/d} } \frac{ \zeta_R ((d-1)/2) }
{\zeta_R (d/2) ^{(d-1)/d} } \frac 1 {N^{1/d}} \right\}\nn
\eeq
where 
\beq
T_0 = \frac{h^2}{2\pi mk} \left(\frac N {{\rm vol} (\cam) 
\zeta_R (d/2)}\right)^{2/d} .\nn
\eeq
The rectangular three dimensional box has been analysed in detail 
in \cite{box} using a density of states approach.
\newpage
{\bf Example 2:} $d$-dimensional isotropic 
harmonic oscillator as confining potential.

Let us now choose the confining potential 
\beq
V(x) = \frac 1 2 \hbar \omega m \sum_{i=1}^d x_i ^2 \nn
\eeq
In this case the coefficients are found by just calculating with 
geometric series \cite{finiten}. The result is $b_0 = (\hbar \omega)^{-d},
$ $b_1 = (\hbar d \omega /2) b_0$
and $j_0 = d$, $j_1 = d-1$ and so
\beq
T_c = T_0 \left\{ 1- \frac{\zeta_R (d-1) }{ 2 \zeta_R (d) ^{(d-1)/d}} 
\frac 1 {N^{1/d}} \right\}\nn
\eeq
with
\beq
T_0 = \hbar \omega \left( \frac N {\zeta_R (d) } \right) ^{1/d}\nn
\eeq
(see also f.e. \cite{mullin}).

{\bf Example 3:} $3$-dimensional anisotropic harmonic osciallator as 
confining potential.

In the anisotropic case the relevant potential reads
\beq
V(x) = \frac 1 2 \hbar m (\omega_1 x_1 ^2 + \omega_2 x_2 ^2 
+\omega_3 x_3 ^2 ). \nn
\eeq
With the same comments as above one gets here $j_0 =3$, $ j_1 =2$, and 
\beq
b_0 = \frac 1 {\hbar ^3  \omega_1 \omega_2 \omega_3}; \qquad
b_1 = \frac 1 {2\hbar^2} \left( \frac 1 {\omega_1 \omega_2} +\frac 1 
{\omega_1 \omega_3} +\frac 1 {\omega_2 \omega_3} \right).\nn
\eeq
For the critical temperature this means 
\beq
T_c = T_0 \left\{ 1-\frac{\zeta_R (2)} {3\zeta_R (3) ^{2/3}} 
\gamma N^{-1/3} \right\}\nn
\eeq
with 
\beq
\gamma = \frac 1 2 (\omega_1 \omega_2 \omega_3 ) ^{2/3} 
\left[ \frac 1 {\omega_1 \omega_2} +\frac 1
{\omega_1 \omega_3} +\frac 1 {\omega_2 \omega_3}
\right]\nn
\eeq
and 
\beq
T_0 = \hbar (\omega_1 \omega_2 \omega_3 ) ^{1/3} 
\left( \frac N {\zeta_R (3) } \right) ^{1/3}.\nn
\eeq
These results have also been obtained using an approach based on the 
Euler-Maclaurin formula \cite{ravndal} and based on a density of states 
approach \cite{martin}. 

Analogous formulas for the $d$-dimensional anisotropic harmonic osciallator
can be found in an easy manner.

{\bf Example 4:} Arbitrary power law confining potential in $3$ dimensions.

Let us finally consider the potential 
\beq
V(x) = \epsilon^p \sum_{i=1}^3 |x_i|^p ,\nn
\eeq
where in order to simplify comparison with the harmonic oscillator 
calculation we use the notation 
\beq
\epsilon = \sqrt{\frac m 2} \hbar^{\frac{2-p}{2p}} \omega ^{\frac{p+2} {2p}}.
\nn
\eeq
Using the results of \cite{resum} 
for the resumed heat-kernel, one gets for this case 
\beq
b_0 = \frac {8 \Gamma^3 (1/p) } {\pi^{3/2} p^3} \frac 1 
  {(\hbar \omega )^{\frac 3 2 (1+2/p)}} \nn
\eeq
and $b_1 = E_0 b_0$, together with 
$j_0 = \frac 3 2 +\frac 3 p$ and $j_1 = j_0 -1$. The resulting critical
temperature for this example is then easily found to be
\beq
T_c = T_0 \left\{ 1-\frac{(E_0/\hbar \omega) \zeta_R (1/2 +3/p) }
{(3/2 +3/p ) \zeta_R (3/2 +3/p ) ^{\frac{p+6}{3(p+2)} }} 
\left(\frac{2\Gamma (1/p) }{\sqrt{\pi} p} \right) ^{\frac{2p}{p+2}}
\frac 1 {N^{\frac{2p}{3(p+2)}}} \right\}.\nn
\eeq
The bulk critical temperature here is 
\beq
T_0 = \hbar \omega \left( \frac{\sqrt{\pi}p}{2\Gamma (1/p)} \right) ^{\frac
{2p} {p+2} }\left( \frac N {\zeta_R (3/2+3/p)}\right) ^{\frac{2p}{3(p+2)}}
\nn
\eeq
(for further bulk properties see \cite{MIT}).

In summary we have shown, that our approach enables us to deal with 
complicated external conditions in an efficient way. 
Of course we do not claim that the results presented in the examples 
cannot be obtained by other means; in fact, most of the results for $T_c$ 
have been obtained differently. However, the unified treatment of 
many different situations seems very elegant to us.
Having at hand this
general approach given an ideal Bose gas under arbitrary external conditions
one only has to determine the leading heat-kernel coefficients 
of the associated Schr\"odinger operator 
(or even only take them from the literature) in order to know the whole
thermodynamical behaviour up to or below the critical temperature. 
Without technical complications our approach includes the effect 
of a finite number of particles.

These ideas are also applicable to the canonical ensemble. Furthermore, 
due to the connections between the microcanonical and the grand canonical 
ensemble \cite{mikro3} 
it seems possible to develop 
this approach also for the microcanonical ensemble.

\vspace{1cm}
\noindent{\bf Acknowledgements}
This investigation has been partly supported by the DFG under contract
number BO1112/4-2.

\end{document}